# A comparative study of the physical properties of layered transition metal nitride halides $MNCl$ ($M = Zr, Hf$): DFT based insights

Shaher Azad[1], B. Rahman Rano[1], Ishtiaque M. Syed[1], S. H. Naqib[2,*]

[1]Department of Physics, University of Dhaka, Dhaka-1000, Bangladesh

[2]Department of Physics, University of Rajshahi, Rajshahi-6205, Bangladesh

*Corresponding author e-mail: salehnaqib@yahoo.com

**Abstract**

$ZrNCl$ and $HfNCl$ belong to a class of layered transition metal nitride halides $MNCl$ ($M = Zr, H$). They are from the space group R-3m (No-166) and crystallize in the rhombohedral structure. Both of these materials have shown promising semiconducting behaviors. Recent studies have shown their versatility as semiconductors and also as superconductors when intercalated with alkaline metals. This paper explores the mechanical, optical and electronic properties of these two semiconducting crystals in depth. A comparative study between the two materials in their elastic constants, anisotropy measures, electronic density of states and band structures, optical spectra has been performed with first principles density functional theory (DFT) based calculations within the local density approximation (with appropriate $U$ for the energy gap calculations in case of *HfNCl*). $HfNCl$ is more machinable than $ZrNCl$ and is relatively softer as indicated by the lower Debye temperature. $ZrNCl$ has stronger layering due to which it exhibits brittle nature. $HfNCl$ has a larger band gap. $ZrNCl$ is a better reflector of ultraviolet radiation. On the other hand $HfNCl$ is a good ultraviolet absorber. Both materials are anisotropic in regards to structure, electronic energy dispersion and optical parameters. Overall, the degree of anisotropy is more prominent in $ZrNCl$ compared to that in $HfNCl$. Possible sectors for applications of *ZrNCl* and *HfNCl* semiconductors are discussed.

**Keywords:** Density functional theory (DFT); Structural properties; Elastic constants; Optoelectronic properties

## 1. Introduction

Semiconductors have dominated the field of electronic devices as their properties can be controlled and tuned through doping. Therefore, it comes as no surprise that a great deal of research has been focused on finding and studying novel semiconductors which have potential applications in next generation electronic devices. Such research is of vital importance to the industry and will help in solving the present day energy crisis. In recent years, layered semiconducting materials have gathered much attention due to their potential use in high performance electronic devices and other related fields. The subjects of our study, $ZrNCl$ and $HfNCl$, belong to a class of layered transition metal nitride halide semiconductors. This class of metal nitride halides, *MNX* (*M* = Ti, Zr, Hf; *X* = Cl, Br, I), was first reported by Juza and Heners in 1964 [1]. $ZrNCl$ and $HfNCl$ have shown promising thermoelectric properties such as substituting the $Zr$ atom with $Hf$ atom in $ZrNCl$ increases the thermal conductivity of the crystal due to a phenomenon called lanthanide contraction and has potential application in creating next generation thermoelectric devices [2,3]. It has been found in mono-layer $HfNCl$ and $ZrNCl$ that even at high temperatures the bonds between the transition metal ion ($Zr, Hf$) and the surrounding nitrogen ($N$) and chlorine ($Cl$) did not break-down, which implies the bonds of the crystal are quite strong [3]. The application of electron doping by intercalation to these materials also generates interesting physical properties such as superconductivity. In particular, intercalation with alkali metal, such as lithium and also sodium and potassium with $\beta$-phases of $ZrNCl$ has been well studied [4–7]. The behavior of the material can change drastically depending on the nature of the intercalating atom. Surprisingly, electron-doped $HfNCl$ shows superconductivity at 25.5 K. This phenomenon has been studied in depth [8–12]. Studies on structural properties have been performed for the two-dimensional layered $Na_xHfNCl$ [13]. Muon spin relaxation $(\mu SR)$ measurements was performed on intercalated crystals particularly $Li_xHfNCl$ and $Li_xZrNCl$ [14,15]. It is noted that $Na_xHfNCl$, $Li_xHfNCl$ and $Li_xZrNCl$ exhibit superconductivity as well. High-pressure experiments reveal that these layered materials possess unusually high electron-phonon coupling constants [16], resulting in relatively high superconducting transition temperature. The effects of intercalation, such as electro-chemical insertion of $Zn$ can go as far as to change the space-group of the crystal system [17]. The critical temperature of superconductivity can be controlled to a large extent by the nature of doping. Reduction doping in $Li_xZrNCl$ has demonstrated an increase in its critical temperature [18]. Electron band structure calculations of the layered transition metal nitride halides have been performed for the 2D case [19]. The Raman spectra of these materials was also analyzed [20]. Electron doped $\beta$-$ZrNCl$ and $\beta$-$HfNCl$ have shown swelling effect [21]. Doping with Magnesium gave rise to a family of novel superconductors [22]. Experimental studies on these classes of materials have also been performed. Of particular note is the study of the effects on the unconventional superconductivity by varying the electron doping concentration [23]. It has also been possible to induce superconductivity to these materials by applying electric field [24]. Recently, they have also been studied as novel 2D semiconductors [25]

Therefore, it is of great interest that a comprehensive characterization of these materials is performed. We have thoroughly reviewed available literature on $ZrNCl$ and $HfNCl$, and to the best of our knowledge, have found no prior studies that explore a number of important physical properties, including elastic, thermo-mechanical, and optical properties of these materials. Analyses of the machinability index, tetragonal shear modulus, acoustic velocities, Kleinman parameter, Cauchy pressure, anisotropy indices etc. are still lacking. Mulliken population analysis and Hirshfeld population analysis also requires exploring. Properties such as Debye temperature, thermal expansion melting temperature, heat capacity etc. have not been investigated thoroughly yet. A deeper understanding of the acoustic, elastic, electronic and optical properties of these materials is crucial to unravel their potential in future applications. In this paper, we have studied all these properties of the layered semiconductors, motivated from the wish to bridge the existing research gaps. We have calculated the bulk elastic properties of the materials and also their optical properties. The band structures and band gaps along with density of states have been revisited. Population analysis and bond analysis have been done for both the compounds. Thermal properties were also evaluated. ELATE plot for both the materials were presented to investigate the elastic anisotropy.

The rest of this paper has been organized as follows: In section 2, we have described briefly the density functional theory (DFT) based computational methodology that we have used for all our calculations. In section 3, we have tabulated and analyzed the obtained results from computations. In section 4, we have summarized the key features and important highlights of our study and have drawn conclusions with relevant discussion.

## 2. Computational Methodology

The properties of $ZrNCl$ and $HfNCl$ have been calculated from first principles calculations using the CAmbridge Serial Total Energy Package (CASTEP) [26]. The various properties of $ZrNCl$ and $HfNCl$ were derived using the DFT formalism as the basis. In DFT, the many-body Schrödinger equation is restated in terms of equivalent non-interacting problem where the system is described in terms of electron density, instead of solving it directly since the system is characterized by the complicated and difficult to compute wave functions. The ground state electron density of the crystal system is obtained by solving the Kohn-Sham equations [27]. Both Generalized Gradient Approximation (GGA-PBE) and Local Density Approximation (LDA) were used as the exchange correlation functionals [28,29]. It is known that GGA tends to overestimate lattice parameters, while LDA sometimes underestimates them. Since the LDA optimized structure gave lattice parameters close to the reference values, all subsequent calculations were performed using that structure for both the materials. The LDA+$U$ method with different Hubbard $U$ values was used only to determine the band gap of $HfNCl$. Since the band gap of $ZrNCl$ calculated from the LDA method agreed well with available literature, the LDA+$U$ method was not used for this compound. Vanderbilt ultra-soft pseudopotential was used during the geometry optimization since it allows for lower basis cut-off, without sacrificing significant computational accuracy [30]. For sampling the Brillouin zones of the compounds under study, a $8 \times 8 \times 8$ k-points mesh was used based on Monkhorst-Pack scheme [31]. To optimize the geometry of the crystal, Broyden–Fletcher–Goldfarb–Shanno (BFGS) algorithm was used [32]. The following electronic orbitals of the atoms: Zr·[$4s^2$ $4p^6$ $4d^2$ $5s^2$], Hf·[$5d^2$ $6s^2$], N·[$2s^2$ $2p^3$] and Cl·[$3s^2$ $3p^5$] were used to perform the pseudo atomic calculations. The plane wave basis set cut-off energy was set to 490 eV to ensure convergence. The geometry was optimized under an energy tolerance of $5.0 \times 10^{-6}$ eV/atom, maximum displacement tolerance of $5.0 \times 10^{-4}$ Å, maximum stress tolerance of 0.02 GPa, and maximum force tolerance of 0.01 eV/Å, with fixed basis set corrections [33]. These tolerance levels have produced good estimates of elastic, structural and electronic band structure properties with an optimum computational time. All the first principle calculations have been performed at the ground states of the compounds. To calculate the single crystal elastic constants $C_{ij}$, the stress-strain method have been employed [34] as contained within CASTEP.

## 3. Results and analysis

### 3.1. Crystal structure

The structures of $ZrNCl$ and $HfNCl$ belonging to the space group $R$-$3m$ (No-166) crystallize in the rhombohedral (trigonal) shape as shown schematically in Fig.1. Each $Zr$ or $Hf$ atom is surrounded by 4 nitrogen atoms and 3 chlorine atoms. The atoms form a layered slab like structure with a van der Waals structural gap between them. Crystallographic parameters optimized under both GGA and LDA for the conventional cell with corresponding data from available literature are presented in Table 1. As expected, the parameters obtained by GGA are higher than those obtained with LDA. It can be seen that our results obtained from first principles calculations are in good agreement with experimental results when local density approximation was adopted [35].

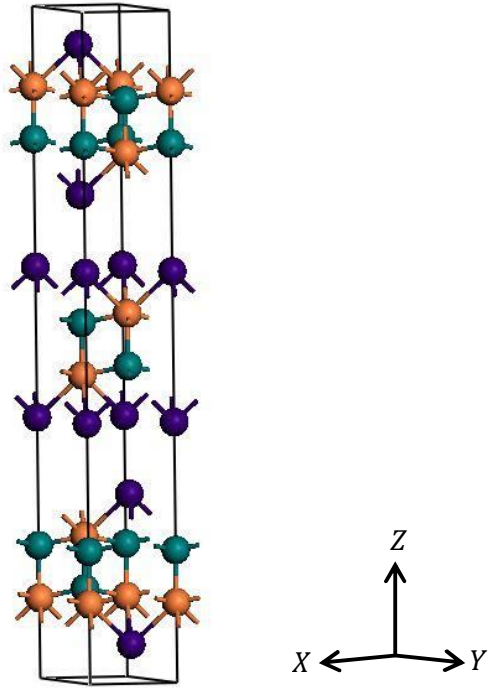

**FIG. 1.** Crystal structure of $ZrNCl$ and $HfNCl$. The orange spheres represent the $Zr$ or $Hf$ atoms, the green spheres represent the $N$ atoms, and the purple spheres represent the $Cl$ atoms.

**TABLE 1:** The lattice parameters and cell volumes of $HfNCl$ and $ZrNCl$ compounds obtained with GGA and LDA. The lattice parameters *a*, *b*, and *c* are in Å and the unit cell volume *(V)* is in Å$^3$.

| Compound | Method | *a* | *b* | *c* | Cell volume (*V*) |
|---|---|---|---|---|---|
| $HfNCl$ | GGA | 3.70 | 3.70 | 31.05 | 368.94 |
|  | LDA | 3.65 | 3.65 | 27.12 | 312.83 |
|  | Reference [35] | 3.58 | 3.58 | 27.62 | 307.01 |
| $ZrNCl$ | GGA | 3.63 | 3.63 | 32.22 | 366.75 |
|  | LDA | 3.56 | 3.56 | 27.33 | 300.67 |
|  | Reference [35] | 3.60 | 3.60 | 27.67 | 311.38 |

Since the values obtained from the LDA calculations were closer to the experimental results than those obtained with GGA, all subsequent calculations were performed using the LDA optimized crystal structure.

*3.2. Mechanical and elastic properties*

The elastic properties of a crystalline material correspond to many of its mechanical, thermal, and lattice dynamical qualities. The elastic properties of $ZrNCl$ and $HfNCl$ need to be investigated to determine and understand the response of these compounds to applied stresses of different nature. Due to their rhombohedral structure, these crystals have 6 independent elastic constants, namely, $C_{11}, C_{12}, C_{13}, C_{14}, C_{33}$ and $C_{44}$. The calculated values of these constants are listed in Table 2. Mechanical stability of the crystals can be investigated from the Born criteria, as modified for rhombohedral crystals [36,37]:

$$C_{11} > |C_{12}|, C_{44} > 0, C_{13}^2 < \frac{1}{2}C_{33}(C_{11} + C_{12}), C_{14}^2 < \frac{1}{2}C_{44}(C_{11} - C_{12}) \equiv C_{44}C_{66} \qquad (1)$$

All of these criteria were satisfied by both of the materials, indicating that they are mechanically stable.

**TABLE 2:** The single crystal elastic constants $C_{ij}$ (GPa), Cauchy pressure $C'' = (C_{12} - C_{44})$(GPa), tetragonal shear modulus $C'$ (GPa), and Kleinman parameter $\zeta$ calculated for $HfNCl$ and $ZrNCl$ compounds.

| Compound | $C_{11}$ | $C_{12}$ | $C_{13}$ | $C_{14}$ | $C_{33}$ | $C_{44}$ | $C''$ | $C'$ | $\zeta$ |
|---|---|---|---|---|---|---|---|---|---|
| *HfNCl* | 205.20 | 128.96 | 57.60 | -16.04 | 54.43 | 46.75 | 82.21 | 38.12 | 0.73 |
| *ZrNCl* | 194.23 | 92.16 | 31.09 | -7.12 | 30.67 | 46.57 | 45.59 | 51.03 | 0.60 |

The resistance to linear compression along [100] and [001] directions are quantified by the elastic constants $C_{11}$ and $C_{33}$, respectively. For both $ZrNCl$ and $HfNCl$, it can be seen that $C_{11}$ is substantially larger than $C_{33}$, which indicates that the bonding strength along [100] direction are much stronger than that along [001] direction. The obtained uniaxial stiffness constants follow the sequence $C_{11} = C_{22} > C_{33}$, which implies that the bonding along *c*-direction is the weakest. This is a consequence of the layered character of $ZrNCl$ and $HfNCl$ and weak interlayer structural coupling. The resistance to shear deformation with respect to a tangential stress applied to the (100) plane in the [010] direction of the crystal is represented by the elastic constant $C_{44}$. The value of $C_{44}$ for $HfNCl$ is much lower than both $C_{11}$ and $C_{33}$, which indicates that the crystal is more easily deformed by a shear in comparison to a stress that is along any of the three crystallographic directions. However for $ZrNCl$, the value of $C_{44}$ is much lower than $C_{11}$ but higher than $C_{33}$ meaning that the crystal is deformed easily by a shear than a stress along the first two crystallographic directions defining the basal plane. For $HfNCl$, $C_{44}$ is higher than $C_{66}$, which indicates that the shear along the (001) plane is easier relative to the shear along the (100) plane. For $ZrNCl$ the opposite is true since the value of $C_{44}$ is lower than $C_{66}$. The value of $C_{44}$ for both the material is close to each other. Since $C_{11} + C_{12} > C_{33}$, bonding in the (001) plane is more rigid than that along the *c*-axis as well as the elastic tensile modulus is higher on the (001) plane than that along the *c*-axis for both the compounds. The tetragonal shear modulus ($C' = (C_{11} - C_{12})/2$), also known as the shear constant, of a crystalline compound is a measure of the crystal's stiffness (the resistance to shear deformation by a shear stress applied in the (110) plane in the [1$\bar{1}$0] direction). The value of $C'$ is related to the degree of dynamic stability of a crystal. A positive value of shear constant of a material indicated the dynamic stability with respect to tetragonal distortions. The value of $C'$ for $ZrNCl$ and $HfNCl$ is 51.03 GPa and 38.12 GPa, respectively. As both the values are positive, the crystals are predicted to be dynamically stable. Since the value of $C'$ for $ZrNCl$ is higher than that of $HfNCl$, it can be implied that $ZrNCl$ is more stable with respect to time varying deformations compared to $HfNCl$.

The Kleinman parameter ($\zeta$), or the internal strain parameter, is a dimensionless elastic indicator that describes the relative position of the cation and anion sublattices under volume conserving distortions. For these distortions atomic positions are not fixed by the crystal symmetry. This parameter measures the relative ease of bond bending to that of bond stretching. A compound's dynamic stability against bending and stretching type distortions is also indicated by this parameter [38]. The Kleinman parameter ($\zeta$), is calculated from the following relation [39]:

$$\zeta = \frac{C_{11} + 8C_{12}}{7C_{11} + 2C_{12}} \qquad (2)$$

The value of the Kleinman parameter generally lies in the range $0 \leq \zeta \leq 1$. The upper and lower limits of $\zeta$ ($\zeta = 1$ and $\zeta = 0$, respectively) are the indicator of the insignificant contribution of bond stretching or contracting to resist the applied stress and the insignificant contribution of bond bending to resist the external stress, respectively. The calculated values of $\zeta$ for $ZrNCl$ and $HfNCl$ are 0.60 and 0.73, respectively. From which we can say that the mechanical strengths in both of these crystals should be dominated by bond stretching or contracting contributions.

The isotropic bulk modulus $(B)$ and shear modulus $(G)$ were estimated by the widely employed VRH (Voigt-Reuss-Hill) method. The Young's modulus $(Y)$, Poisson's ratio $(\nu)$ and hardness $(H)$ of the crystals are calculated using the following relations [40–42]:

$$B_H = \frac{B_V + B_R}{2} \tag{3}$$

$$G_H = \frac{G_V + G_R}{2} \tag{4}$$

$$Y = \frac{9BG}{(3B + G)} \tag{5}$$

$$\nu = \frac{(3B - 2G)}{2(3B + G)} \tag{6}$$

$$H = \frac{(1 - 2\nu)Y}{6(1 + \nu)} \tag{7}$$

Isotropic bulk modulus and shear modulus are rough estimates of the overall bonding strength of a material. The bulk modulus is a measure of resistance of change in volume when a uniform pressure is applied and the shear modulus is a measure of resistance to deformations that are reversible when a shearing stress is present. For both $ZrNCl$ and $HfNCl$, greater value of $B$ compared to $G$ (Table 3) indicates that the mechanical strengths of these compounds are limited by shear deformation. The resistance against tension or compression along the length of a material can be determined from the Young's modulus [43]. With increasing Young's modulus the covalent nature of a material increases [44]. The Young's modulus also provides a measure of thermal shock resistance. A material is said to be stiff if the value of its Young's modulus is large. The calculated value of Young's modulus $(Y)$ is comparatively large for both the compounds. Therefore, both of them are expected to be reasonably stiff. The values are close to each other indicating that they have almost the same degree of stiffness. The lattice thermal conductivity and Young's modulus of a material are connected by the following relation [45]:

$$K_L \approx \sqrt{Y} \tag{8}$$

Thus, the compounds under investigation should have notable phonon thermal conductivity.

The $G/B$ ratio, also known as Pugh's ratio, is a measure of brittleness/ductility as proposed by Pugh [46]. A solid is ductile if the value of the ratio is lower than 0.57, otherwise it is predicted to be brittle in nature. In our case, for $ZrNCl$ the ratio is 0.71 which indicates brittle nature. In contrast, For $HfNCl$, it is 0.44 which indicates ductile nature.

The Poisson's ratio is a critical indicator for observing a number of mechanical and bonding properties of a crystal. This ratio of a material can take values between -1.0 to 0.50 [40]. If the Poisson's ratio $v$ for a material is $v \leq 0.33$, it should exhibit brittle nature; otherwise it should be ductile in nature [47]. It can be seen from Table 3 that the calculated value of the ratio $v$ for $ZrNCl$ is 0.21 and for $HfNCl$ it is 0.31. Both indicate that the solids are brittle in nature. However, since the value of the Poisson's ratio for $HfNCl$ is almost near the margin of brittleness/ductility and Pugh's ratio suggests that $HfNCl$ should be ductile, it might be possible that the Poisson's ratio by itself is not a definitive differentiator of the plastic nature of this crystal. The lower and upper limits of the Poisson's ratio for central force solids are 0.25 and 0.50 respectively [48]. We can see that according to those criteria $HfNCl$ is a central force solid and the interatomic forces of $ZrNCl$ are non-central in nature. The value of Poisson's ratio also indicates the existence of covalent and ionic bonding in a compound. For covalent bonding $v \sim 0.10$ and for ionic bonding $v \sim 0.25$ [49]. We see that the calculated values for both the solids imply the presence of significant ionic contribution. Another measure for brittle/ductile nature of a solid is the Cauchy pressure ($C'' = C_{12} - C_{44}$). If the Cauchy pressure is positive the material is said to be ductile otherwise it should be brittle [50]. From Table 2 we see that both materials have positive Cauchy pressure which indicates ductility in both solids. Since, corrections due to many body interactions among atoms and electron gas are not taken into consideration while calculating the Cauchy pressure, it can give misleading results. Such ambiguity has been observed in other layered compounds. Positive Cauchy pressure suggests that in addition to ionic bonding, some metallic bonding is also present in both crystals according to the Pettifor's rule [51].

Machinability index, an important engineering indicator, defines the ease with which a material can be cut or shaped while providing a satisfactory surface finish. The machinability index is given by the expression:

$$\mu_M = \frac{B_H}{C_{44}} \qquad (9)$$

The machinability of a solid is influenced by a number of factors such as the inherent characteristics of the working compounds, cutting tool geometry, tool composition, cutting fluid, cutting conditions, type of cutting and machine tool rigidity [38]. Machinability index is a valuable measure in today's industry as it defines the optimum machine utilization, temperature, cutting forces and power strain. The machinability index is also used as a measurement of plasticity and dry lubricity of solids [52–54]. The machinability index of $ZrNCl$ is 1.20 and the machinability index of $HfNCl$ is 1.71. This implies that $HfNCl$ is more machinable than $ZrNCl$. This may be due to the fact that the Pugh's ratio for the former compound is lower than the later. The machinability index of $HfNCl$ is comparable to some of the industrially prominent layered MAX and MAB phase compounds [57-60].

To understand the plastic and elastic properties of a material better, it is also important to study the hardness value. The calculated value of hardness of $ZrNCl$ is 7.51 GPa and for $HfNCl$ it is 4.49 GPa. This suggests that $ZrNCl$ is much harder than $HfNCl$, and this is consistent with the obtained values of the machinability index. The values of the elastic moduli, ratios and other indicators are summarized in Table 3A and Table 3B.

**TABLE 3A:** The calculated isotropic bulk modulus $B$ (in GPa), shear modulus $G$ (in GPa), and Young's modulus $Y$ (in GPa) of $HfNCl$ and $ZrNCl$ compounds deduced by Voigt, Reuss, and Hill (VRH) approximations.

| Compound | $B_V$ | $B_R$ | $B_H$ | $G_V$ | $G_R$ | $G_H$ | $Y$ |
|---|---|---|---|---|---|---|---|
| $HfNCl$ | 105.91 | 54.34 | 80.12 | 41.03 | 30.06 | 35.55 | 92.91 |
| $ZrNCl$ | 80.86 | 30.67 | 55.77 | 46.49 | 32.31 | 39.40 | 95.67 |

**TABLE 3B:** The calculated Pugh's indicator $G/B$, Machinability index $\mu_M = B_H/C_{44}$, Poisson's ratio $\nu$, and hardness $H$ of $HfNCl$ and $ZrNCl$ compound deduced by Voigt, Reuss, and Hill (VRH) approximations.

| Compound | $G_V/G_R$ | $B_V/B_R$ | $G/B$ | $\mu_M$ | $\nu$ | $H$ |
|---|---|---|---|---|---|---|
| $HfNCl$ | 1.37 | 1.95 | 0.44 | 1.71 | 0.31 | 4.49 |
| $ZrNCl$ | 1.44 | 2.64 | 0.71 | 1.20 | 0.21 | 7.51 |

*3.3. Elastic anisotropy*

Almost all known crystals are elastically anisotropic. Anisotropy of the elastic parameters describes the directional dependence of mechanical properties. So, it is important to have a good knowledge about the elastic/mechanical anisotropy because it is closely related to formation of micro cracks in crystals, motion of cracks, development of plastic deformations etc. The measure of the degree of anisotropy in the bonding between atoms in different planes is provided by the shear anisotropic factors. There are three different shear anisotropic factors that can be used to explain the anisotropic nature of the elastic parameters of a crystal [61].

The shear anisotropic factor for the {100} shear planes between the [011] and [010] directions is:

$$A_1 = \frac{4C_{44}}{C_{11} + C_{33} - 2C_{13}} \tag{10}$$

The shear anisotropic factor for the {010} shear planes between the [101] and [001] directions is:

$$A_2 = \frac{4C_{55}}{C_{22} + C_{33} - 2C_{23}} \tag{11}$$

The shear anisotropic factor for the {001} shear planes between the [110] and [010] directions is:

$$A_3 = \frac{4C_{66}}{C_{11} + C_{22} - 2C_{12}} \qquad (12)$$

The shear anisotropic factors obtained from calculations are recorded in Table 4. Crystals that have isotropic nature in the bonds existing between the planes have unit values of the shear anisotropic factors i.e., $A_1 = A_2 = A_3 = 1$, and any other value implies that there is anisotropy present in the crystal. From Table 4, we see that both $HfNCl$ and $ZrNCl$ are anisotropic in nature. The nature of anisotropy in both materials follows a similar pattern. In both materials $A_1 = A_2$ suggesting that the anisotropy between those planes are the same. It is interesting to note that $A_3 = 1$ for both crystals, implying isotropy in that plane. Since both crystals have similar structure and geometry, the results are consistent.

**TABLE 4:** Shear anisotropic factors $(A_1, A_2, A_3)$, anisotropy in shear $A^G$, anisotropy in compressibility $A^B$, the universal anisotropy index $A^U$, equivalent Zener anisotropy measure $A^{eq}$, and universal log-Euclidean index $A^L$ calculated for $HfNCl$ and $ZrNCl$.

| Compound | $A_1$ | $A_2$ | $A_3$ | $A^B$ | $A^G$ | $A^U$ | $A^{eq}$ | $C_{44}^R$ | $C_{44}^V$ | $A^L$ |
|---|---|---|---|---|---|---|---|---|---|---|
| HfNCl | 1.29 | 1.29 | 1 | 0.32 | 0.15 | 2.77 | 4.06 | 14.29 | 14.72 | 0.67 |
| ZrNCl | 1.14 | 1.14 | 1 | 0.45 | 0.18 | 3.83 | 4.99 | 16.09 | 16.19 | 0.97 |

The universal log Euclidean index $A^L$ is given by the equation [62,63]:

$$A^L = \sqrt{\left[\ln\left(\frac{B^V}{B^R}\right)\right]^2 + 5\left[\ln\left(\frac{C_{44}^V}{C_{44}^R}\right)\right]^2} \qquad (13)$$

where $C_{44}^V$ and $C_{44}^R$ are the Voigt and Reuss approximated values of $C_{44}$. These are calculated from the relations [62]:

$$C_{44}^R = \frac{5}{3}\left\{\frac{C_{44}(C_{11} - C_{12})}{3(C_{11} - C_{12}) + 4C_{44}}\right\} \qquad (14)$$

and

$$C_{44}^V = C_{44}^R + \frac{3}{5}\left\{\frac{(C_{11} - C_{12} - 2C_{44})^2}{3(C_{11} - C_{12}) + 4C_{44}}\right\} \qquad (15)$$

The expression of $A^L$ is valid regardless of crystal symmetry. For a perfectly isotropic crystal, $A^L = 0$. The values of $A^L$ has a range between 0 and 10.26. Although almost 90% of crystalline materials have $A^L < 1$. Compounds having higher value of $A^L$ indicate that they are strongly layered and compounds having small values have non-layered structures. From the obtained values of $A^L$ it can be observed that $ZrNCl$ has stronger layering than $HfNCl$. This might be another possible reason of the greater machinability of $HfNCl$ between the two crystals. The value of $A^L$ is closely related with the universal anisotropy index $A^U$. The universal anisotropy index $A^U$, anisotropy in shear $A^G$, anisotropy in compressibility $A^B$, and equivalent Zener anisotropy measure $A^{eq}$ of any compound can be calculated using the relations [62,64,65]:

$$A^U = 5\frac{G_V}{G_R} + \frac{B_V}{B_R} - 6 \qquad (16)$$

$$A^B = \frac{B_V - B_R}{B_V + B_R} \qquad (17)$$

$$A^G = \frac{G_V - G_R}{2G_H} \qquad (18)$$

$$A^{eq} = \left(1 + \frac{5}{12}A^U\right) + \sqrt{\left(1 + \frac{5}{12}A^U\right)^2 - 1} \qquad (19)$$

Elastic isotropy is indicated by $A^B = 0$ and $A^G = 0$, and the maximum possible anisotropy is given by $A^B = A^G = 1$. From Table 4 we see that both of the crystals are anisotropic in bulk compressibility and shear. The value of anisotropy in compressibility and shear is larger for $ZrNCl$. Both the compounds are highly anisotropic in compressibility. The magnitude of the anisotropy in shear is smaller compared to the anisotropy in compressibility. The universal anisotropy factor $A^U$, first introduced by Ranganathan and Ostoja-Starzewski [64], can be applied to any crystal symmetry. $A^U = 0$ indicates elastic isotropy and any other positive value means anisotropic nature. The equivalent Zener anisotropy index $A^{eq}$ also indicates elastic isotropy for $A^{eq} = 0$ and anisotropy for any other value. We can see that both of the crystals are anisotropic in nature and $ZrNCl$ is more anisotropic than $HfNCl$ in every anisotropy measure.

To visualize the anisotropy on bulk elastic behavior in three dimensions (3D), we have shown the ELATE generated plots in Figures 2 and 3 [66]. Solids that are isotropic in nature should exhibit spherical shapes in three dimensional Young's modulus, Poisson's ratio, shear modulus and compressibility plots. Any other shape would indicate deviation from the isotropic nature.

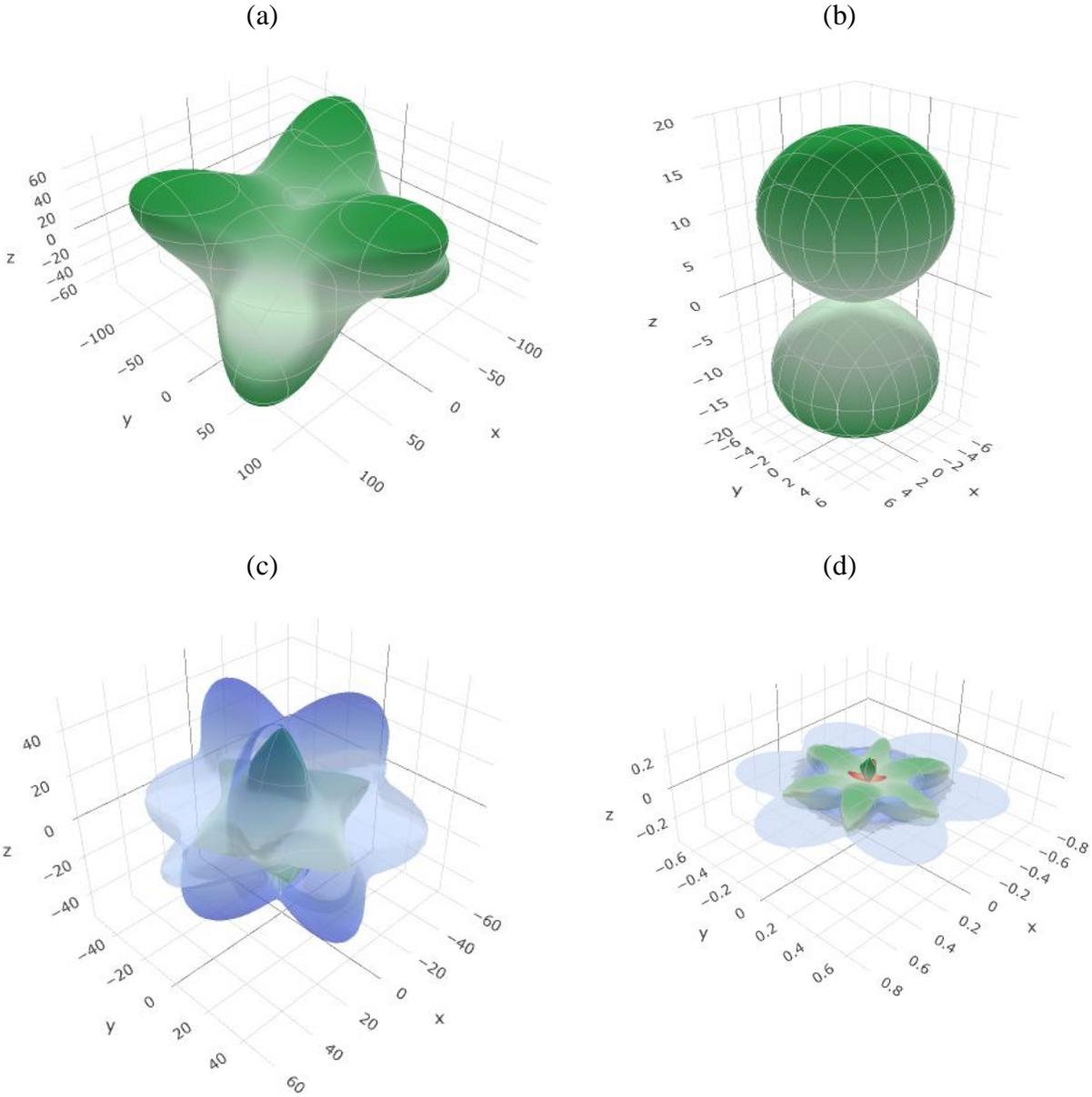

**FIG. 2.** Three dimensional directional dependences of (a) Young's modulus (b) compressibility (c) shear modulus and (d) Poisson's ratio for $HfNCl$.

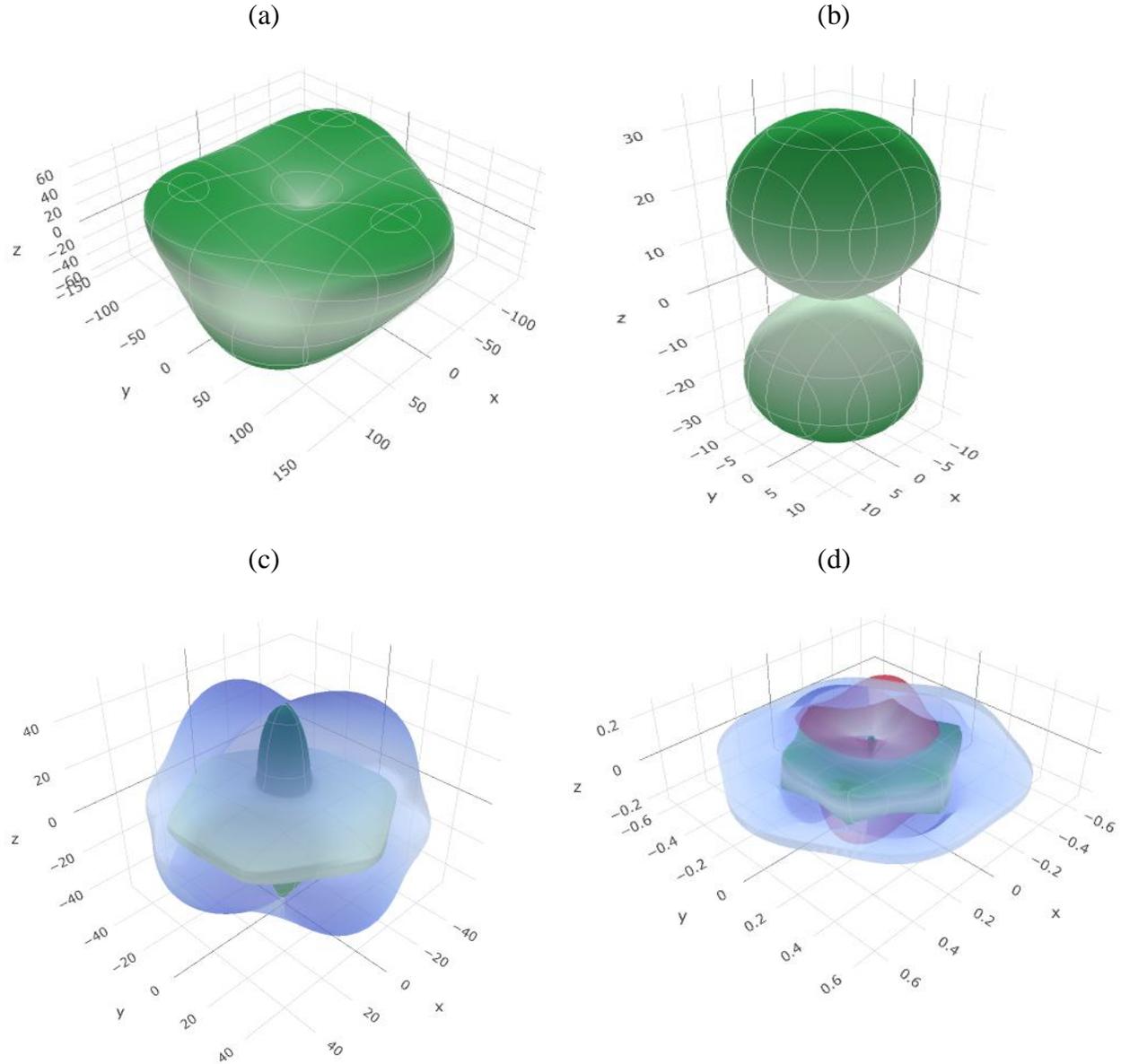

**FIG. 3.** Three dimensional directional dependences of (a) Young's modulus (b) compressibility (c) shear modulus and (d) Poisson's ratio for $ZrNCl$.

These plots indicate that both the compounds are highly anisotropic. These anisotropies originate from the strongly layered structure of the crystals. Extreme anisotropy in the compressibility (inversely proportional to the bulk modulus) is striking. This indicates that inserting atomic layers in both $HfNCl$ and $ZrNCl$ is very easy on top of the *ab*-plane. Such features are seen in many of the recently discovered van der Waals solids [67].

*3.4. Acoustic velocities*

To understand a material's mechanical properties, electrical and thermal conductivities, the velocity of sound in a material is a parameter of significance. Crystals with high sound velocity, such as diamond, are known as the best heat conductors at room temperature. In this section, the phase velocity of longitudinal and transverse modes of $HfNCl$ and $ZrNCl$ have been studied. Bulk modulus, $B$ and shear modulus, $G$, can be used to determine the transverse and longitudinal velocities of sound in a crystalline compound using the relations [68]:

$$V_t = \sqrt{\frac{G}{\rho}} \quad (20)$$

and

$$V_l = \sqrt{\frac{B + \frac{4G}{3}}{\rho}} \quad (21)$$

Here $\rho$ is the mass density of the compound. From these equations we can see that the velocities are strongly dependent on the density of the solid and elastic moduli. A material with zero shear modulus cannot support the transverse mode of sound. The average sound velocity is given by:

$$V_a = \left[\frac{1}{3}\left(\frac{2}{V_t^3} + \frac{1}{V_l^3}\right)\right]^{-\frac{1}{3}} \quad (22)$$

The acoustic impedance of materials is useful because it regulates the transfer of acoustic energy between two media. The study of acoustic impedance of materials has become an important tool for transducer design, noise reduction and many other acoustic applications. The amount of acoustic energy transmitted and reflected at the interface of two materials when sound is transmitted from one to another is determined by the difference in acoustic impedances. When the difference in impedance is large most of the sound is reflected. This results in the loss of transmitted signal. On the other hand, if the two impedances are equal or close to one another most of the energy is transmitted. The acoustic impedance of $HfNCl$ and $ZrNCl$ have been calculated using the following equation [38,69]:

$$Z = \sqrt{\rho G} \quad (23)$$

The acoustic impedance has a unit of Rayl. $1\ Rayl = m^4/kg.s = kgm^{-2}s^{-1} = 1\ Nsm^3$. We can see from the equation that a material with high density and high shear modulus has high acoustic impedance.

The intensity of sound radiation $I$ is related to the radiation factor $\sqrt{\frac{G}{\rho^3}}$ as [38,69]:

$$I \approx \sqrt{G/\rho^3} \qquad (24)$$

This factor is important for the selection of materials for sound board design. The calculated values of all these parameters for both $HfNCl$ and $ZrNCl$ are shown in Table 5.

**TABLE 5:** Density $\rho(kg/m^3)$, transverse sound velocity $v_t(ms^{-1})$, longitudinal sound velocity $v_l(ms^{-1})$, average sound wave velocity $v_a(ms^{-1})$, acoustic impedance $Z$ (Rayl), and radiation factor $\sqrt{G/\rho^3}$ ($m^4/kg.s$) of $HfNCl$ and $ZrNCl$.

| Compound | $\rho$ | $v_t$ | $v_l$ | $v_a$ | $Z(\times 10^6)$ | $\sqrt{G/\rho^3}$ |
|---|---|---|---|---|---|---|
| HfNCl | 7259.67 | 2212.84 | 4191.13 | 2473.81 | 16.06 | 0.31 |
| ZrNCl | 4661.7055 | 2907.15 | 4819.91 | 3214.36 | 13.55 | 0.62 |

From Table 5 we can see that the value of the transverse, longitudinal and average velocity of $ZrNCl$ is greater than that of $HfNCl$. The acoustic impedance of $HfNCl$ is higher. The acoustic radiation factor of $ZrNCl$ is 0.62, almost double of that of $HfNCl$, so it is a much better material for sound board designing.

*3.5. Thermal properties*

The Debye temperature of solids can be calculated from their elastic constants. At the Debye temperature the phonon wavelength becomes close to the unit cell length. The Debye temperature is important because it is related to a number of thermal and electrical properties of a crystal such as specific heat, melting temperature, thermal expansion coefficient, thermoelectricity, resistivity etc. The superconducting transition temperatures of phonon mediated superconductors are linked closely with the Debye temperature [70,71]. The Debye frequency $\omega_d$ is the highest allowed phonon frequency in compound according the Debye model. The Debye temperature $\Theta_d$ is related to the Debye frequency via the formula:

$$\Theta_d = \hbar\omega_d/k_B \qquad (25)$$

The Debye temperature is also related to the hardness of a material. A material with low Debye temperature tends to be a soft material. Materials with higher Debye temperature show greater hardness, greater interatomic bonding strength, higher melting temperature etc. The Debye temperature can be calculated from the average sound velocities as follows [72]:

$$\Theta_D = \frac{h}{k_B}\left(\frac{3n}{4\pi V_o}\right)^{\frac{1}{3}} v_a \qquad (26)$$

where $h$ is the Planck's constant, $k_B$ is the Boltzmann's constant, $n$ is the number of atoms in the molecule, $V_o$ is the volume of the optimized unit cell, $v_a$ is The average sound velocity calculated in the previous section. The calculated value of the Debye temperatures of $HfNCl$ and $ZrNCl$ crystals are disclosed in Table 6.

The melting temperature $T_m$ is an important parameter that is related to the thermal expansion coefficient, bonding energy and elastic constants. It is crucial to study this parameter to know about the thermal limits of a material for high temperature application. Materials whose melting temperature is high tend to have stronger atomic bonding and lower thermal expansion coefficient [69]. The elastic constants of $HfNCl$ and $ZrNCl$ can be used to estimate the melting temperature using the empirical relation [73]:

$$T_m = 354K + (4.5K/GPa)\left(\frac{2C_{11} + C_{33}}{3}\right) \pm 300K \tag{27}$$

The calculated values are given in Table 6. Since there is a substantial uncertainty of $\pm 300\ K$ this formula is only a rough measure of the melting temperature.

Thermal expansion coefficient $(\alpha)$ is an intrinsic parameter of solids. A number of physical properties, such as thermal conductivity, specific heat are correlated with the thermal expansion coefficient. The thermal expansion coefficient of a solid can be evaluated from the following equation [74]:

$$\alpha = \frac{1.6 \times 10^{-3}}{G} \tag{28}$$

where $G$ is the shear modulus of the solid. The calculated value of the thermal expansion coefficients of $HfNCl$ and $ZrNCl$ crystals are also shown in Table 6.

The dominant phonon wavelength of a compound, $\lambda_{dom}$, defines the wavelength at which the phonon energy distribution curve reaches its maximum. Compounds that typically have large shear modulus and low density demonstrate large values of the dominant phonon mode. The dominant phonon mode at 300 K is given by the formula [75]:

$$\lambda_{dom} = \frac{12.566 v_a}{T} \times 10^{-12} \tag{29}$$

The heat capacity per unit volume $(\rho C_P)$ of a material defines the change in thermal energy per unit volume in a material per degree Kelvin change in the temperature. The heat capacity of a material per unit volume in the high temperature limit, where the Dulong-Petit law holds, can be calculated from [74]:

$$\rho C_P = \frac{3k_B}{\Omega} \tag{30}$$

where, $N = 1/\Omega$ is the number of atoms per unit volume. The calculated values of dominant phonon mode at 300 K and heat capacity per unit volume for $HfNCl$ and $ZrNCl$ crystals are shown below in Table 6.

**TABLE 6:** The Debye temperature $\Theta_D$ (K), thermal expansion coefficient α ($K^{-1}$), wavelength of the dominant phonon mode at 300 K, $\lambda_{dom}$ (m), melting temperature $T_m(K)$ and heat capacity per unit volume $\rho C_P (J/m^3.K)$ of $HfNCl$ and $ZrNCl$.

| Compound | $\Theta_D$ | α (× $10^{-5}$) | $\lambda_{dom}$ (× $10^{-12}$) | $T_m$ | $\rho C_P$ (× $10^{-6}$) |
|---|---|---|---|---|---|
| HfNCl | 284.34 | 4.50 | 102.7 | 1051.25±300 | 2.38 |
| ZrNCl | 374.37 | 4.06 | 134.6 | 982.69±300 | 2.48 |

From Table 6 we can see that the Debye temperature of $HfNCl$ is lower than that of $ZrNCl$. This suggests that $HfNCl$ is relatively softer than $ZrNCl$. This is consistent with the hardness values of both materials shown in Table 3. This is also consistent with the measures of brittle/ductile nature and the machinability indices calculated before. Although the melting temperature of $HfNCl$ is found to be higher, it is inconclusive since there is the uncertainty of $\pm 300\,K$. The wavelength of the dominant phonon mode and heat capacity per unit volume of $ZrNCl$ is higher, consistent with the calculated Debye temperature. The thermal expansion coefficient of $ZrNCl$ is lower than $HfNCl$, which is also consistent with the values of their respective Debye temperatures.

*3.6. Electronic properties*

*3.6.1. Electronic band structure*

The electronic band structure contains a lot of information about any compound. The overlap of the electronic wave functions of every atom in a crystal creates the band structure which is shown in the band structure plot (Fig. 4 and Fig. 5) for both $HfNCl$ and $ZrNCl$. They are shown along paths that pass through specific high symmetry points of the Brillouin zone for the ground state optimized crystal structure. The plots show the valence band and the bottom of the conduction band within an energy range of -5eV to 5eV for both the materials. The horizontal line in the middle at 0 eV is the Fermi level, $E_F$. Both of the materials are semiconducting in nature and have indirect band gaps. The band gap was found to be 2.238 eV and 1.759 eV respectively for $HfNCl$ and $ZrNCl$ which agree well with the existing known values [76,77]. The LDA+$U$ method was used in determining the band gap of $HfNCl$, since the obtained value by the LDA alone was found to be 1.83 eV, which underestimated the reference value. This is due to the fact that the electronic correlation in $HfNCl$ is stronger than that in $ZrNCl$, and this must be taken into account when performing the band gap calculations. The band gap was tuned by gradually increasing the Hubbard $U$ value. As the $U$ value was increased, the value of the band gap also increased. A Hubbard $U$ value of 7.08 corresponded to a band gap of 2.238 eV, which agreed well with existing literature values. Since the value of the band gap for $ZrNCl$ obtained with LDA matched well with available literature, the LDA+$U$ method was not required for this particular compound.

The band structures illustrate several interesting features. For $HfNCl$ the bands along H-K are more dispersive than the bands along A-H or K-Γ. The top of the valence band is located at the Γ point. Similarly for $ZrNCl$ the bands are more densely packed along the A-H, K-Γ and Γ-M lines and are dispersive along the H-K regions. Both materials have similar band structure though the conduction band of $HfNCl$ is more densely packed than the conduction band of $ZrNCl$. The differences in the degree of electronic energy dispersion along different directions are indicative of the electronic anisotropy. For example, nearly flat $E(k)$ curves along the *c*-direction (H-K and M-L segments within the Brillouin zone) implies high electronic effective mass. The degree of dispersion is higher in the bottom region of the conduction band than that in the top part of the valence band. These are the sections which dominate the charge transport properties of a semiconductor.

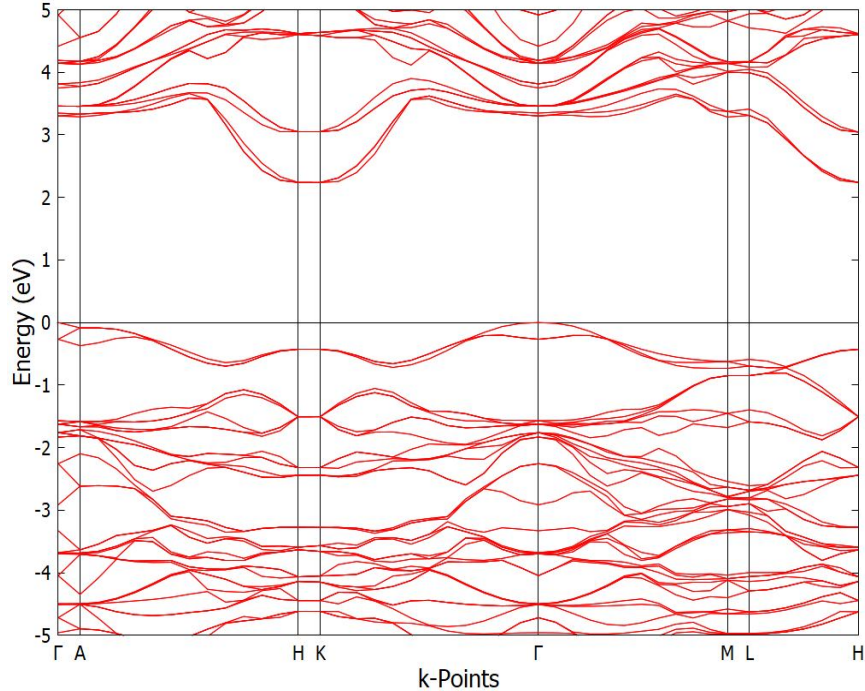

**FIG. 4.** The electronic band structure of $HfNCl$ along points of high symmetry in the Brillouin zone.

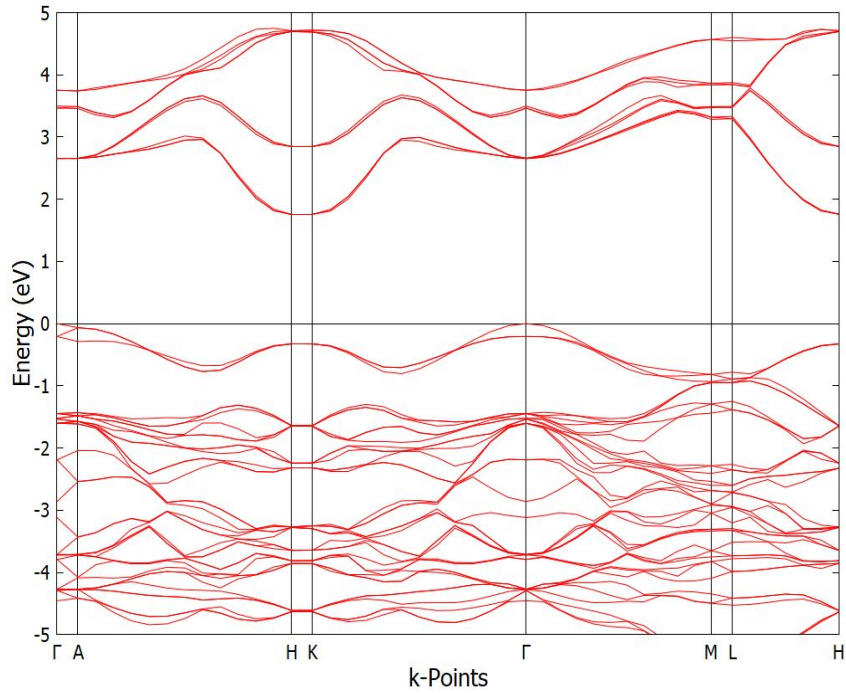

**FIG. 5.** The electronic band structure of $ZrNCl$ along points of high symmetry in the Brillouin zone.

*3.6.2. Electronic density of states*

The calculated total density of states (TDOS) and the partial density of states (PDOS) of individual atoms of $HfNCl$ and $ZrNCl$ are shown in Fig. 6 and Fig. 7, respectively.

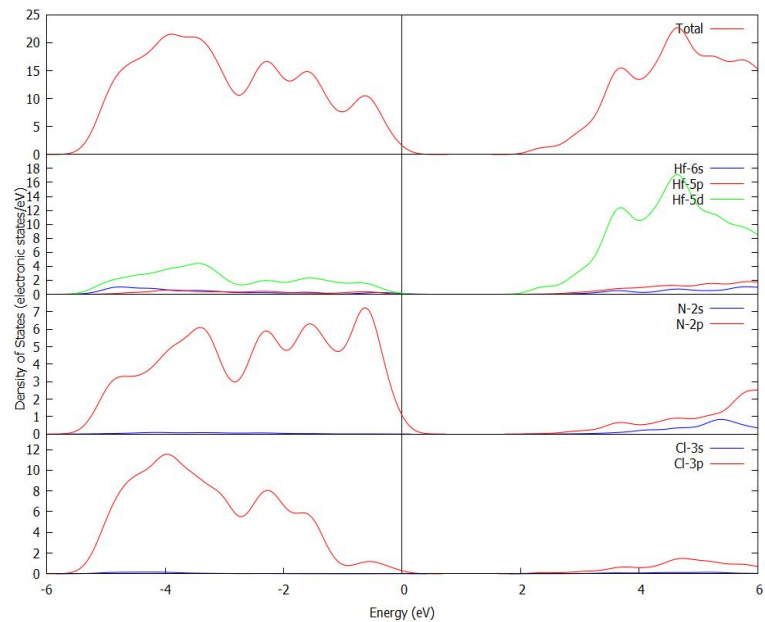

**FIG. 6.** The total and partial density of states of $HfNCl$.

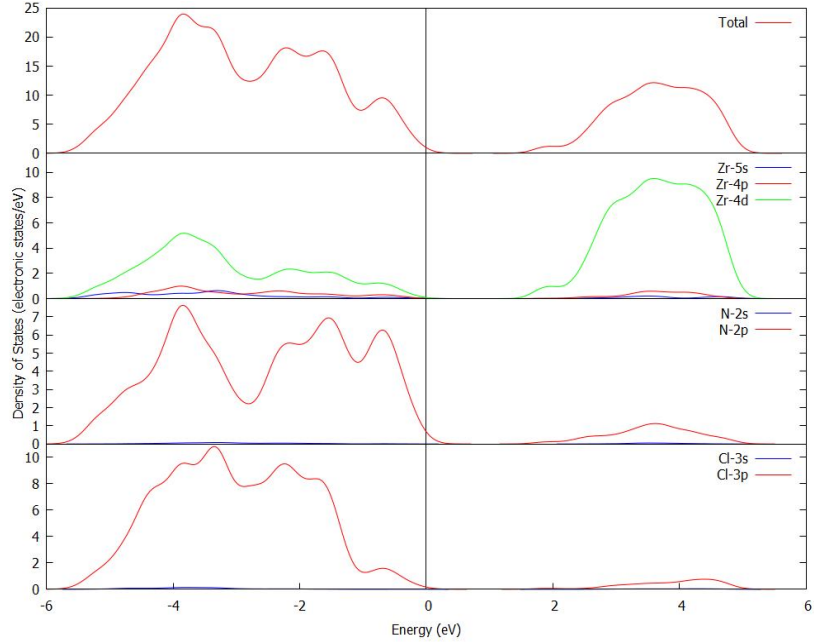

**FIG. 7.** The total and partial density of states of $ZrNCl$.

From the TDOS of $HfNCl$ it can be seen that there is a gap above the Fermi level at 0 eV. This gap corresponds to the band gap of the crystal. The density of states is high before and after the gap. From the PDOS of individual atoms it can be seen that most of the contribution to the TDOS comes from Hf atoms above the Fermi level. While below the Fermi level, the contribution mainly comes from the N and Cl atoms and Hf atoms have insignificant contribution in that region. From the PDOS of the Hf atoms it is evident that the Hf-5d orbitals mainly contribute to the density of states and the Hf-5p and Hf-6s atomic orbitals have very little contribution. Similarly for $ZrNCl$, it can be seen that the gap above the Fermi level is also present and it exactly matches with the value of the obtained band gap. Above the Fermi level, the Zr atoms contribute while below the Fermi level the N and Cl atoms have contribution. From the PDOS of the Zr atoms it is also clear that the Zr-4d orbitals are the main contributors of the density of states and the Zr-4p and Zr-5s atomic orbitals have little contribution. For both materials, it can be observed from the PDOS of N and Cl atoms that N-2p and Cl-3p mainly contribute to the density of states and the N-2s and Cl-3s have almost no or insignificant contribution. Thus the conduction band of $HfNCl$ is mainly a mix of Hf-5d, N-2p and Cl-3p orbitals and similarly the conduction band of $ZrNCl$ is a mixture of Zr-4d, N-2p and Cl-3p orbitals. The TDOS of $HfNCl$ is larger than the TDOS of $ZrNCl$ above the Fermi level. For both compounds there is large hybridization among the Zr-4d/Hf-5d and Na-2p and Cl-3p electronic states near the top of the valence band. This is indicative of covalent bonding between the constituent atoms. The strengths of these covalent bonds largely determine the hardness, Debye temperature and melting point of the compounds under study.

*3.7. Bond population analysis*

To understand the bonding nature of the crystals better, the Mulliken Population Analysis has been used [78]. The effective valence charge (EVC) is the difference between formal ionic charge and the calculated Mulliken charge. Non-zero EVC for all of the species indicate covalent nature in their chemical bonding. The charge spilling parameter represents the quantity of missing valence charges. The Mulliken charge and the Hirshfeld charge for both $HfNCl$ and $ZrNCl$ compounds are calculated and shown below in Table 7.

**TABLE 7:** Charge spilling parameter (%), orbital charges (electron), atomic Mulliken charge (electron), and Hirshfeld charge (electron) for $HfNCl$ and $ZrNCl$.

| Compound | Atoms | No of ions | Charge Spilling (%) | s | p | d | Total | Mulliken Charge | EVC | Hirshfeld Charge |
|---|---|---|---|---|---|---|---|---|---|---|
| *HfNCl* | Hf | 6 | 0.35 | 0.43 | 0.17 | 2.37 | 2.96 | 1.04 | 2.96 | 0.43 |
|  | N | 6 |  | 1.71 | 4.06 | 0.00 | 5.77 | -0.77 | 3.77 | -0.38 |
|  | Cl | 6 |  | 1.94 | 5.33 | 0.oo | 7.27 | -0.27 | 1.27 | -0.06 |
| *ZrNCl* | Zr | 6 | 0.20 | 2.32 | 6.32 | 2.30 | 10.94 | 1.06 | 2.94 | 0.43 |
|  | N | 6 |  | 1.68 | 4.12 | 0 | 5.80 | -0.80 | 3.80 | -0.40 |
|  | Cl | 6 |  | 1.93 | 5.32 | 0 | 7.26 | -0.26 | 1.26 | -0.03 |

The low values of charge spilling parameters indicate a good representation of electronic bonds where charge spillings for $HfNCl$ and $ZrNCl$ are only 0.35% and 0.20%, respectively. For $HfNCl$ the Mulliken atomic charges for Hf, N and Cl atoms are 1.04, -0.77 and -0.27, respectively. This indicates that electronic charge is transferred from Hf atoms to N and Cl atoms. Similarly for $ZrNCl$, the values of the Mulliken charges indicate that electronic charge is transferred from Zr atoms to N and Cl atoms. For both crystals the values of the Mulliken charges indicate that covalent bond is present in the crystals with significant ionic contributions. The values are similar for both crystals indicating that both of the compounds contain similar type of bondings. The Hirshfeld charges for both materials are lesser than the Mulliken charges. The Hirshfeld population analysis often gives better results than Mullikan population analysis [79]. The Hirshfeld population analysis agrees qualitatively with the Mulliken population analysis in the matter of electronic charge transfer. The degree of ionic contribution is smaller in the case of Hirshfeld population analysis. The bond overlap population was also calculated and is summarized below in Table 8.

**TABLE 8:** Calculated bond overlap populations and bond lengths (Å) for $HfNCl$ and $ZrNCl$.

| Compound | Bond | Bond Number | Population | Bond Length (This work) | Bond Length (Experimental [80]) |
|---|---|---|---|---|---|
| HfNCl | N-Hf | 6 | 1.47 | 2.15 | - |
| | N-Hf | 6 | 0.54 | 2.21 | - |
| | N-N | 3 | -0.19 | 2.74 | - |
| | Cl-Hf | 6 | 0.77 | 2.75 | - |
| ZrNCl | N-Zr | 6 | 1.11 | 2.11 | 2.11 |
| | N-Zr | 6 | 0.42 | 2.14 | 2.14 |
| | N-N | 3 | -0.22 | 2.67 | 2.71 |
| | Cl-Zr | 6 | 0.73 | 2.71 | 2.77 |

Table 8 shows that the N-Hf bond has a larger population than the N-Zr bond. In fact between the similar bonds in both crystals $HfNCl$ has a larger population in all cases. On the other hand, all the bond lengths are smaller in $ZrNCl$. This is responsible for higher hardness of this compound in comparison with $HfNCl$. Moreover, the bond populations are highest for the N-Hf and N-Zr bonds in *HfNCl* and *ZrNCl*, respectively. These bonds are prime contributors in overall mechanical strengths of the compounds under study. The overlap populations which are close to zero indicate insignificant interaction between the atoms involved. The estimated bond lengths for *HfNCl* show excellent agreement with those found from experiments [74].

*3.8. Optical properties*

The optical properties of a crystalline solid describe its behavior in response to incident electromagnetic radiation and are therefore important to study to find its usefulness in optoelectronic applications. The optical parameters investigated herein include (a) the complex dielectric function $\varepsilon(\omega)$, (b) refractive index $\eta(\omega)$, (c) conductivity $\sigma(\omega)$, (d) reflectivity $R(\omega)$, (e) absorption coefficient $\alpha(\omega)$ and (f) loss function $L(\omega)$. All these parameters have been determined for both $HfNCl$ and $ZrNCl$ for the polarization directions [100] and [001] of incident electric field vector. The imaginary part of the dielectric function is evaluated from the CASTEP supported formula [81]:

$$\varepsilon_2(\omega) = \frac{2e^2\pi}{\Omega\varepsilon_0} \sum_{k,v,c} |\langle\psi_k^c|\boldsymbol{u}\cdot\boldsymbol{r}|\psi_k^v\rangle|^2 \delta(E_k^c - E_k^v - E) \tag{31}$$

where, $\omega$ is the angular frequency of the incident electromagnetic wave (photon), $\Omega$ is the volume of the unit cell and $e$ is the charge of an electron. In this equation, $\boldsymbol{u}$ is the polarization vector of the incident electric field. $\psi_k^c$ and $\psi_k^v$ are the conduction and valence band wave functions at a wave vector *k*, respectively. The delta function implements conservation of energy and momentum during the optical transition. The Kramers-Kronig transformations can then be used to find out the real part of the dielectric function. Other optical parameters can be obtained from the dielectric functions, using standard formalisms [82].

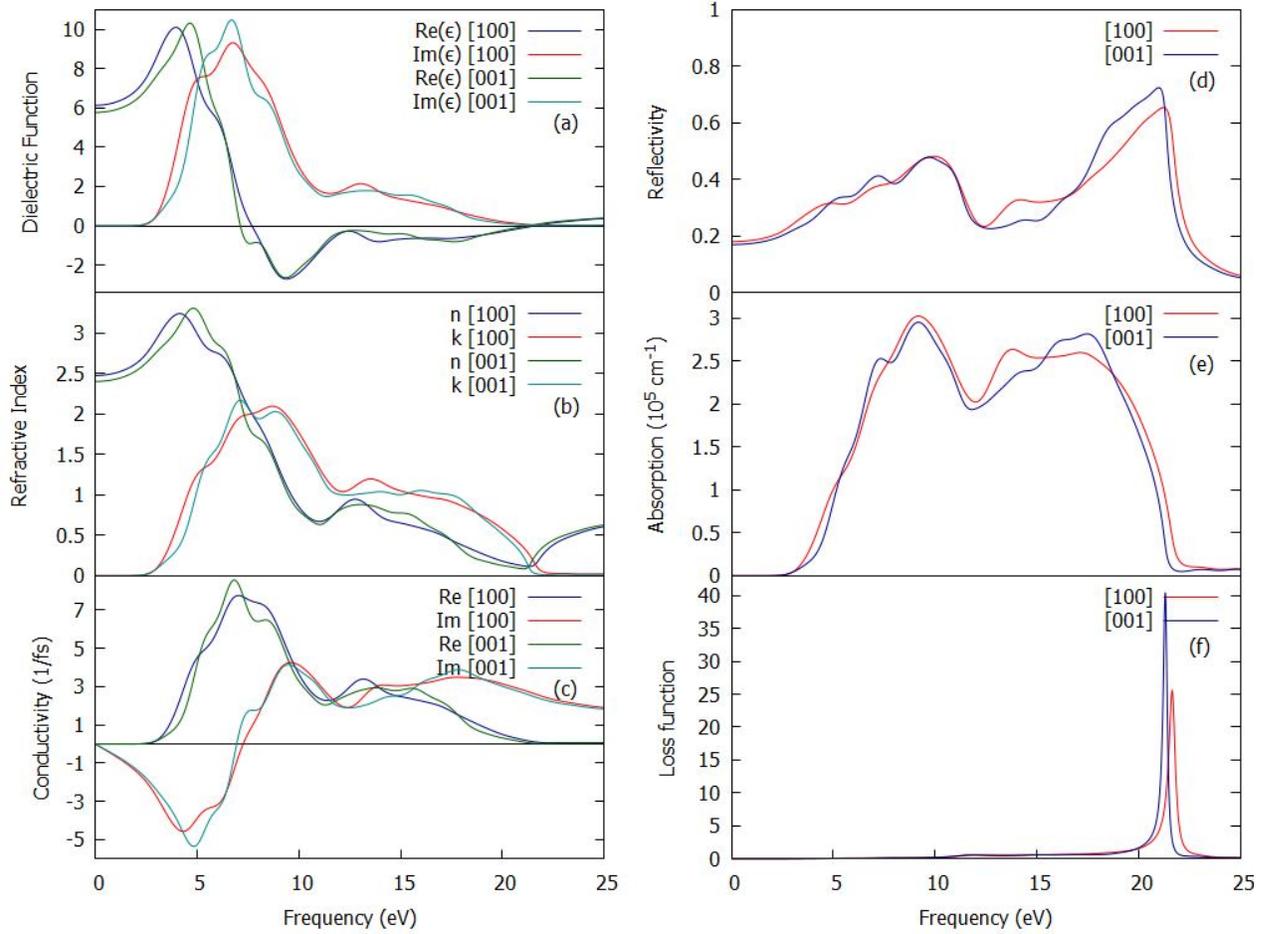

**FIG. 8.** The frequency dependent (a) dielectric function (real & imaginary parts), (b) refractive index (real & imaginary parts), (c) optical conductivity (real & imaginary parts), (d) reflectivity, (e) absorption coefficient, and (f) loss function of $HfNCl$ (with polarization directions along [100] & [001]).

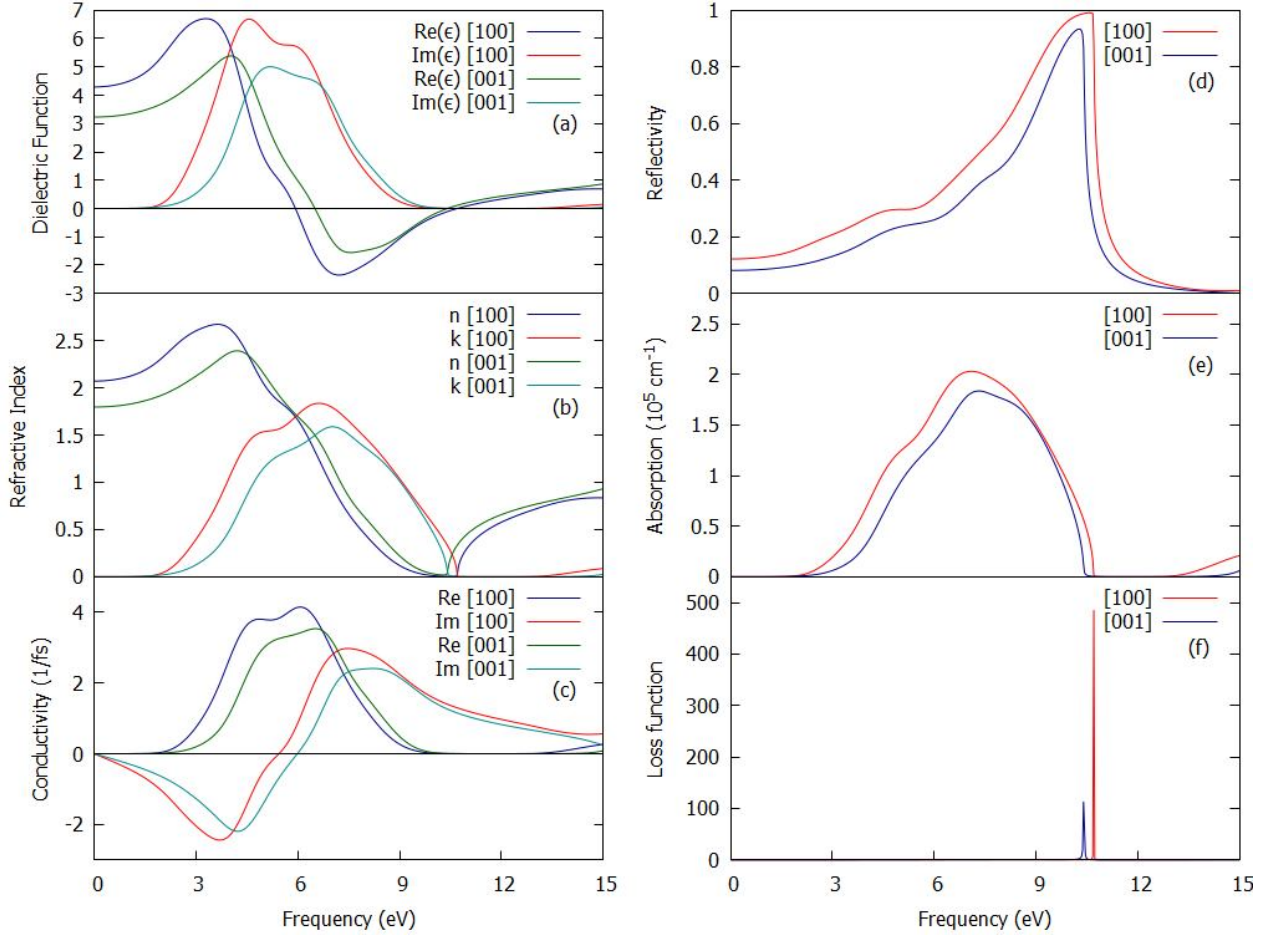

**FIG. 9.** The frequency dependent (a) dielectric function (real & imaginary parts), (b) refractive index (real & imaginary parts), (c) optical conductivity (real & imaginary parts), (d) reflectivity, (e) absorption coefficient, and (f) loss function of $ZrNCl$ (with polarization directions along [001] & [001]).

Figures 8 and 9 show the graphs of the calculated optical spectra for $HfNCl$ and $ZrNCl$ for the [100] and [001] polarization directions of the incident electric field, respectively. For $HfNCl$, from Fig. 8(a) and we can see that the dielectric function peaks near 5 eV and approaches zero after 22 eV, close to the plasma frequency of this compound. This is true for both of the polarizations. For $ZrNCl$, the peak is also near 5 eV, but the plasma frequency is near 11 eV as seen from Fig. 9(a). This difference in the plasma frequency indicates a significant difference in the charge dynamics between the two compounds. Above the plasma frequency, a material becomes nearly transparent to incident electromagnetic radiation. The real part of the dielectric function measures polarization. This is negative in the region of 8 eV to 22 eV for $HfNCl$. It also has a negative value for $ZrNCl$ in the range of 7 eV to 11 eV. The negative value means that the polarization diminishes the applied electric field in response to it. Both of the polarization directions for both crystals follow the same trend for this parameter. Although the negative region spans a greater range for $HfNCl$, $ZrNCl$ reaches the lowest value in terms of amplitude, meaning it diminishes an applied electric field more strongly than $HfNCl$ in that particular region. The imaginary of the dielectric function signifies loss of energy of the electromagnetic radiation within the compounds.

The ratio of the speed of light in a medium to that of free space is represented by the real part of the refractive index. The imaginary part is the extinction coefficient. The absorption coefficient is related to this parameter. For $HfNCl$, the real part is particularly high in the infrared to visible region and peaks near 5 eV then gradually drop to zero near its plasma frequency ~ 22 eV. Similarly for $ZrNCl$, it also drops to zero near 11 eV, while peaking near 4 eV. The large values of the refractive index near the visible region indicate that the crystals under investigations will be a good candidate for optical display devices. The optical conductivity of $HfNCl$ has a greater range of 3 eV to 20 eV, whilst $ZrNCl$ conducts moderately in the 4 eV to 8 eV region, indicating that *HfNCl* is the better optical conductor out of the two crystals.

From Fig. 8(d), it is seen that the reflectivity for $HfNCl$ rises in a gradual manner from the visible to near ultraviolet region. It decreases near 10 eV then gradually rises up to 60%-70% in the upper ultraviolet region. $ZrNCl$ shows a different behavior as can be seen from Fig. 9(d). The reflectivity keeps rising till it reaches a maximum value of 95%-100% near the plasma frequency, dropping sharply after this frequency. This high value of the reflectivity suggests that $ZrNCl$ is a better ultraviolet reflector than $HfNCl$ and has potential use as an efficient ultraviolet reflector. Interestingly, for both the compounds, the reflectivity is quite low in the visible region of the optical spectra. Thus, these compounds can be used as anti-reflection coating.

The absorption coefficients for both the materials follow a similar pattern. Fig. 8(e) and Fig. 9(e) depict the absorption spectra of $HfNCl$ and $ZrNCl$, respectively. The value of the absorption coefficient is an important quantity which determines how efficiently a material absorbs photons of certain energy. It is seen that $HfNCl$ absorbs quiet well in the 7 eV to 20 eV region and $ZrNCl$ absorbs quiet well in the 6 eV to 9 eV region. *HfNCl* has a broader absorption band and can be used as an ultraviolet absorber.

The loss function of $HfNCl$ has a peak near 22 eV, and for $ZrNCl$ it peaks near 11 eV. This corresponds to the plasmon energy of the materials. This is consistent with the drop of reflectivity and absorption function near this energy. This also suggests the transparency of the material above the plasma frequency. The intensity of the peak is greater in the [001] polarization direction for $HfNCl$. However, the loss function of $ZrNCl$ has greater intensity in the [100] polarization direction.

Both the materials exhibit optical anisotropy. *ZrNCl* is more optically anisotropic than *HfNCl*. The degree of anisotropy is most prominent in the loss function spectra. All the features of energy dependent optical parameters show semiconducting behavior in complete agreement with the electronic band structure calculations.

## 4. Conclusions

Mechanical, electronic, thermal and optical properties of layered semiconductors $ZrNCl$ and $HfNCl$ have been investigated in detail using the DFT formalism. Some of the physical properties are explored for the first time. The anisotropic natures of the materials are seen from the elastic, electronic and optical properties. The ELATE generated plots also demonstrate the anisotropic nature of both materials in a graphical manner. It has been found that $ZrNCl$ is more anisotropic than $HfNCl$. $ZrNCl$ possesses a stronger layered structure. $HfNCl$ is fairly machinable, soft and should possess high level of dry lubricity. The mechanical features of $ZrNCl$ and $HfNCl$ are consistent with the thermal properties. The band gaps of both the compounds show promise to be used in the photovoltaic sector. Due to layered structure, intercalation is easy in both the compounds. This suggests that band gap engineering is a viable option for these materials with suitable atomic insertions. Due to low value of the out-of-the plane compressibility, moderate uniaxial pressure along the *c*-direction can also alter the electronic band structure significantly. The electron dispersion curves show large anisotropy. Bond population analysis suggests the covalent nature of bonding between the atoms in both crystals with notable ionic contributions. The compounds are also potential candidates for optical display systems due to having large values of refractive index in the visible region. $ZrNCl$ is a better reflector of ultraviolet radiation of the two. On the other hand, $HfNCl$ is the better absorber of ultraviolet light. Both the compounds reflect visible light poorly and can be used as antireflection coating.

To summarize, a number of hitherto unexplored properties of $ZrNCl$ and $HfNCl$ compounds have been investigated in this work. The compounds possess interesting thermo-mechanical and optical features which can have relevance to applications. We hope that this study will inspire the researchers to explore various physical properties of these compounds in greater detail.

## Acknowledgements

S. A. acknowledges the Research Fellowship from Semiconductor Technology Research Centre, University of Dhaka, Bangladesh. S. H. N. acknowledges the research grant (1151/5/52/RU/Science-07/19-20) from the Faculty of Science, University of Rajshahi, Bangladesh, which partly supported this work.

## Data availability

The data sets generated and/or analyzed in this study are available from the corresponding author on reasonable request.

## Author contributions

S. A. performed the theoretical calculations, contributed to the analysis and draft manuscript writing. B. R. R. performed the theoretical calculations, contributed to the analysis, and contributed to manuscript writing. I. M. S. supervised the project and contributed to finalizing the manuscript. S. H. N. supervised the project, analyzed the results and finalized the manuscript. All the authors reviewed the manuscript.

**Competing Interests**

The authors declare no competing interests.